\documentclass[lettersize,journal]{IEEEtran}
\usepackage{amsmath,amsfonts}
\usepackage{algorithmic}
\usepackage{algorithm}
\usepackage{array}
\usepackage[caption=false,font=normalsize,labelfont=sf,textfont=sf]{subfig}
\usepackage{textcomp}
\usepackage{stfloats}
\usepackage{url}
\usepackage{verbatim}
\usepackage{graphicx}
\usepackage{cite}
\usepackage[numbers]{natbib}
\usepackage{booktabs}
\usepackage{color}
\usepackage{xcolor}

\usepackage{color,colortbl,soulutf8}

\usepackage{ifthen}
\newif\ifshort
\shorttrue   
\ifshort
	\newcommand{\isShort}{true}
\else
	\newcommand{\isShort}{false}
\fi
\newcommand{\shortVer}[1]{\ifthenelse{\equal{\isShort}{true}}{{#1}}{}}
\newcommand{\longVer}[1]{\ifthenelse{\equal{\isShort}{false}}{{#1}}{}}

\ifshort
\else
\fi

\newif\ifcomment
\commenttrue
\ifcomment
\newcommand{\fg}[1]{{\bf\textcolor{red}{FG: #1}}}

\newcommand{\clau}[1]{\textcolor{purple}{#1}}
\newcommand{\edited}[1]{\textcolor{black}{#1}}
\newcommand{\andrea}[1]{\textcolor{teal}{#1}}

\else
\newcommand{\sz}[1]{}
\newcommand{\fg}[1]{}

\newcommand{\clau}[1]{{#1}}
\newcommand{\edited}[1]{{#1}}
\newcommand{\andrea}[1]{{#1}}
\fi

\newcommand{\torevise}[1]{\textcolor{black}{#1}}

\newcommand{\newtext}[1]{\textcolor{black}{#1}}
\newcommand{\new}[1]{\textcolor{black}{#1}}
\newcommand{\moved}[1]{\textcolor{black}{#1}}

\usepackage{navigator}

\usepackage{tikz,xcolor}

\definecolor{lime}{HTML}{A6CE39}
\DeclareRobustCommand{\orcidicon}{
	\begin{tikzpicture}
	\draw[lime, fill=lime] (0,0) 
	circle [radius=0.16] 
	node[white] {{\fontfamily{qag}\selectfont \tiny ID}};
	\draw[white, fill=white] (-0.0775,0.1) 
	circle [radius=0.005];
	\end{tikzpicture}
	\hspace{-3mm}
}
 
\foreach \x in {A, ..., Z}{%
	\expandafter\xdef\csname orcid\x\endcsname{\noexpand\urllink{https://orcid.org/\csname orcidauthor\x\endcsname}{\noexpand\orcidicon}}

}

\renewenvironment{IEEEbiography}[1]
  {\IEEEbiographynophoto{#1}}
  {\endIEEEbiographynophoto}
  
\begin{document}

\title{Inequalities in Computational Thinking among Incoming Students in a STEM Chilean University}

\author{Felipe González-Pizarro\orcidB{}, Claudia López\orcidA{}, Andrea Vásquez\orcidC{}, Carlos Castro\orcidD{}
\thanks{}}


\markboth{}%
{}

\maketitle

\begin{abstract}
While computational thinking arises as an essential skill worldwide, formal primary and secondary education in Latin America rarely incorporates mechanisms to develop it in their curricula. \newtext{The extent to which students in the region acquire computational thinking skills remains largely unknown. To start addressing this void, this paper presents findings from a cross-sectional study that characterizes the computational thinking abilities of incoming students at a Chilean university with a strong emphasis on STEM disciplines. }Based on more than 500 responses, this study provides evidence of significant inequalities in computational thinking across gender, type of school (private or no), and prior programming knowledge. 
The discussion offers insights into how these disparities relate to contextual factors \new{of the country, such as a highly socio-economically segregated educational system, public policies focused mainly on technology access, and heavy reliance on voluntary initiatives to develop computational thinking.} The findings can enlighten upcoming research endeavors and formulate strategies to create a more equitable field for students entering STEM degrees in nations facing similar circumstances.

\end{abstract}


\section{Introduction}

\IEEEPARstart{C}{omputational} thinking refers to the processes required to formulate a problem and express its solution \newtext{so} that an information-processing agent—human or machine—can carry it out effectively~\cite{wing_2006}. It \newtext{relates to} the skills used when generating a computational solution~\cite{wing_2006}. It includes strategies such as logically organizing and analyzing data, representing data through abstractions, automating solutions through algorithmic thinking, \newtext{and} constructing iteratively by debugging previous solutions, among others~\cite{barr2011computational}. 
Its proponents argue that computational thinking is an ability that everyone should acquire to thrive in a society where computer science is ubiquitous~\cite{Monteiro_Leite_2021}. Researchers also \newtext{assert} that it is 
crucial to develop it in primary and secondary education~\cite{wing_2006,barr2011computational,enriquez_2016}. 
 
Australia, \edited{South Korea, Finland, and other 
countries in the Asian Pacific and Europe} have modified their school curricula~\cite{Brackmann_2017_development,heintz_2016,bocconi2016developing,so2020computational} to develop computational thinking \newtext{abilities} and use programming to boost \newtext{these} skills~\cite {moreno_2017_oncomputational}. Other initiatives worldwide have fostered the development of computational thinking in schools~\cite{Lye_2014,heintz_2016} by allowing students to earn academic credits and 
skip introductory computer science courses at university~\cite{Ericson_2014_Measuring}.

However, these trends are still developing in Latin America \new{and the level to which students in the region acquire computational thinking skills remains largely unknown}. There \newtext{have yet to be substantial} changes in curriculum design to introduce computational thinking in the region, even though several efforts are leading in that direction in Argentina, Brazil, and Uruguay~\cite{Pereiro2022}. In Chile, schools \newtext{can} offer computational thinking as an optional course in secondary education \cite{Simmonds2021}. 
Up until now, most initiatives in Latin America are 
isolated attempts by foundations, universities, and schools to promote computational thinking by offering\new{, for example,} programming workshops~\cite{lacoa_2016}. While these actions have attracted significant interest~\cite{Roig-Vila_Moreno-Isac_2020}, the effectiveness of this kind of interventions varies considerably according to a recent meta-review~\cite{merino2022computational}. 
 
This work seeks to provide evidence of the development of computational thinking skills among students who have finished high school in a Latin American country where computational thinking is not a compulsory subject. \new{The first goal is} to understand if students have otherwise developed their computational thinking. \new{The second goal is to assess whether} there are inequalities in the achievement of these skills across \new{gender and type of school (private or not).} 

\new{Prior work has reported mixed evidence regarding a gender gap in computational thinking. There seem to be gender inequalities in computational thinking that broaden as time goes by during primary and secondary school \cite{roman2017cognitive}; however, such inequalities are not observed among students who undertook certain types of training \cite{hutchins2017role,Atmatzidou_2016} and those enrolled in first-year engineering courses \cite{diaz2020computational,ata2020analysis}.} 

\new{Besides, related} \moved{research hints that not only gender but also other students' characteristics may be associated with inequalities in computational thinking. Studies support that racial structures and socioeconomic stratification play a role in the participation and development of related subjects, such as programming, computing, and sciences. Therefore, there is a need to deepen the research on these matters in relationship to computational thinking.}

Evidence has shown that the marginalization of certain social groups relates to the disparities in other science, technology, engineering, and mathematics (STEM) subjects. Working towards equality in these subjects early on is expected to result in more people enrolling in STEM degrees, including various groups who can contribute diverse cultural assets to these fields. Their inclusion has multiple benefits. For example, it can lead to new solutions for problems that affect marginalized communities \cite{lewis201916,shah2020racial}. \new{This study seeks to take initial steps in examining whether such disparities in computational thinking exist and how they may develop in a Latin American country where training these skills is optional.}

This \new{article delves into an assessment} of the computational thinking exhibited by 549 students who were recently admitted to undergraduate programs at a Chilean university exclusively dedicated to STEM disciplines. The study's findings reveal \new{high levels of computational thinking among these students, yet they also unveil substantial disparities based on gender and type of school.}

\moved{Before enrolling in this STEM-focused university, male-identifying students exhibited higher computational thinking scores than their female counterparts. Furthermore, incoming students who had received their high school education in private institutions outperformed those who had attended public or subsidized schools. Moreover, possessing prior programming knowledge was positively associated with elevated levels of computational thinking. Importantly, these observed effects remained statistically significant even after accounting for other relevant variables.}
\new{The authors discuss how the results can inform future research and help shape strategies aimed at fostering fairer conditions for students entering STEM programs in countries with similar contextual factors.}

The rest of this paper \new{is organized as follows. Section \ref{sec:relatedWork} summarizes related work and states the research questions}. 
The next sections present a cross-sectional study of the computational thinking skills of students recently accepted into a STEM-focused university in Chile.
Section \ref{sec:researchmethods} describes the research method. Section \ref{sec:analysissurvey} explains the results.  
Section \ref{sec:discussion} discusses the findings. Lastly, section \ref{sec:conclusion} provide conclusions.

\section{Related Work and Research Questions}
\label{sec:relatedWork}

\newtext{This section outlines the current state of the art and practice regarding computational thinking in Latin America and Chile, aiming to establish the contextual background for this work. Subsequently, the researchers undertake a comprehensive review of various inequalities that could impact the development of computational thinking, both in general and specifically within the region where they conducted their study.} \new{At the end of each subsection, the authors articulate the research question that will be explored in this study.}

\newtext{\subsection{Computational Thinking in Latin America and Chile}}
\label{sec:latam}

\newtext{Experts widely recognize the growing significance of computational thinking skills in effectively navigating the digital landscape. Research has found a strong correlation between these skills and academic performance~\cite{lei2020computational}. 
\new{Efforts aimed at nurturing computational thinking have proven advantageous not only for students in STEM fields but also for those pursuing studies in the humanities and social sciences}~\cite{garcia2018exploring}. 
As a result, the development of computational thinking has gained substantial importance within educational systems across the globe~\cite{Brackmann_2017_development,heintz_2016,bocconi2016developing,so2020computational}.}

\newtext{The rising demand for developing computational thinking has not gone unnoticed in Latin America, which has seen an increase in actions designed to foster these skills in various student populations. A systematic literature review reveals that academic research and computational thinking efforts have grown in Latin America since 2013 compared to little to no activity earlier~\cite{quiroz2021integration}. Over the last ten years, the number of initiatives has rapidly increased. Brazil is the leader in the region regarding policies and non-governmental programs, while Chile and Uruguay are the most active in research~\cite{quiroz2021integration}.}

\newtext{Most computational thinking initiatives in Latin America are workshops and extracurricular activities ~\cite{Pereiro2022,brackmann2016computational}. 
Even though Argentina, Brazil, and Uruguay are heading toward curricula reforms in this regard~\cite{Pereiro2022}, they still need to implement them. Meanwhile, students have had access to numerous voluntary experiences related to computational thinking \cite{Pereiro2022}. This scenario of multiple independent initiatives has brought new challenges to the region. There need to be more teachers who are proficient with digital tools, and, like the rest of the world, there need to be better conceptualizations of what computational thinking entails in public policy. So far, current autonomous efforts are inconsistent with one another \cite{quiroz2021integration}.} 

\newtext{Similarly, in Chile, the initiatives primarily consist of extracurricular activities or workshops offered by upper-level educational institutions and non-governmental programs. For example, several universities offer volunteer programming workshops to primary or secondary school students,\footnote{Programming workshops at Universidad Técnica Federico Santa Maria. Retrieved on July 10, 2023 from: https://ocilabs.cl/} \footnote{Programming workshop at Universidad de Chile. Retrieved on July 10, 2023 from: https://comunicaciones.dcc.uchile.cl/news/691-taller-de-programacion-para-escolares-aprendamos-a-programar-de-manera-ludica/} and non-governmental organizations run similar programs, such as ``Code your ideas'' (``Programa Tus Ideas''\footnote{Website of the ``Code your ideas'' program. Retrieved on July 10, 2023 from \url{https://programatusideas.cl/}} in Spanish).}

\newtext{The Chilean Digital Agenda 2020 declares the population's need for digital education \cite{AgendaDigital_2020}. \new{Nevertheless, government initiatives primarily focus on supplying students with computers to address this concern \cite{fiscarelli2018informe}.} \moved{The only survey related to the use of technology by students in the country, which last took place in 2013, showed that only 1.8\% of students were at an advanced level when using information and communication technologies to carry out tasks related to learning and knowledge \cite{MinistryofEducation_2014}.} Few programs are related to computational thinking, programming, or computing science. The most recent one, the National Plan of Digital Languages\footnote{Description of the National Plan of Digital Languages. Retrieved on July 10, 2023 from https://sitios.mineduc.cl/lenguajesdigitales/que-es-el-plan.html}, focuses on teachers rather than students. It aims to promote the integration of computational thinking in schools by training teachers, enabling them to introduce concepts and skills related to computational thinking. Additionally, schools can offer computational thinking as an optional course in secondary education \cite{Simmonds2021}.} 

\new{Therefore, the predominant volunteer-driven nature of most initiatives directed at school students in Chile and Latin America prompts the question of whether students who have not undergone compulsory computational thinking training have developed these skills.} \moved{To the best of the authors' knowledge, computational thinking levels among school students in Chile and other countries in Latin America have yet to be uncovered.} \new{Thus, the first research question is:}

\begin{itemize}
    \item \new{\textbf{RQ-1: To what extent have computational thinking skills developed among students within an educational system that does not mandate training in these skills?}}
\end{itemize}

\newtext{\subsection{Inequalities in Computational Thinking and STEM}}
\label{sec:inequalities}
In addition to the issues of implementing computational thinking in the educational systems, there is also the challenge of inequality. 
\new{In this subsection, the authors revise prior work related to disparities in the development of computational thinking skills and other STEM-related subjects.} 


\newtext{Previous studies have identified gaps across gender and school grade (or age).} \citet{roman2017cognitive} investigated 1251 Spanish students attending public and private schools, ranging from grades 5 to 10. They observed noteworthy disparities between genders solely in grades 7 to 10, where male students outperformed their female counterparts in the assessment. This divergence became more pronounced in the higher grades, indicating an escalation of the gender gap as educational levels advanced \cite{roman2017cognitive}. 

\newtext{Other studies have confirmed this pattern.} A research project in the UK with 119 upper-secondary students observed that women achieved lower computational thinking performance than men \cite{Lai_2022}. \new{Another research study identified a gender disparity within a sample of 153 high school students in Singapore (124 males and 29 females). The majority (60\%) of male students were assessed as possessing high or very high levels of computational thinking skills, whereas only 7\% of female students exhibited similar proficiency \cite{chan2021assessing}. Besides,} a study in Hong Kong with 13.600 primary school children showed that the students' computational thinking abilities increased with age and that boys performed slightly better than girls \cite{Kong_Lai_2022}. \newtext{In the USA, a project found a gender gap even at much younger ages. In an eight-week robotics and programming intervention with 45 children (from kindergarten to second grade), boys performed better than girls on complex programming tasks but not simpler ones \cite{sullivan2016girls}}.

\newtext{Some authors examined the relationship between gender, training, and computational thinking skills.} A study in the UK showed that the association between programming background and computational thinking performance is significant for men but not women \cite{Lai_2022}. In Greece, a study that involved 164 students from 15 to 18 years old in educational robotics observed that students, regardless of age or gender, could achieve the same level of computational thinking after training. \newtext{Nevertheless,} female students needed more training to achieve the same skill level as their male peers \cite{Atmatzidou_2016}.

However, not all evidence supports the existence of a gender gap in computational thinking skills. A study with 40 high school students conducting a Scratch project in the USA found no significant difference in the grades achieved by women and men but a significant gender gap in confidence levels \cite{hutchins2017role}. In the same country, researchers reported no significant effect of gender in computational thinking across 62 first-year engineering students \cite{diaz2020computational}. Similarly, a study with 112 first-year engineering students in Turkey observed the absence of a gender gap in \newtext{the same subject \cite{ata2020analysis}. Thus, more research is needed to understand better under what circumstances gender inequalities in computational thinking arise and fade.}

\moved{Overall, the evidence shows a complex picture of gender differences in computational thinking. There seems to be a gender gap in both primary \cite{Kong_Lai_2022,sullivan2016girls} and secondary education \cite{Lai_2022,roman2017cognitive}, broadening as time goes by \cite{roman2017cognitive}. On the other hand, there is no discernible gender gap among high school students who have received training in conducting computational thinking projects  \cite{hutchins2017role,Atmatzidou_2016}. Previous research also suggests that gender disparities are absent among students who are enrolled in first-year engineering courses \cite{diaz2020computational,ata2020analysis}.}

\moved{A few hypotheses could explain these results. First, the gender disparities may disappear after reaching a turning point at some moment in the last years of high school. Alternatively, engineering undergraduates are somewhat different than broader student populations because women who can enroll in engineering careers have developed their computational thinking skills similarly to their male counterparts. This study seeks to shed light on this gap in the literature by examining the computational thinking of students admitted to enrolling in STEM degrees right before they start taking courses.} \new{Therefore, the second research question is:}

\begin{itemize}
    \item \new{\textbf{RQ-2: Do gender disparities exist in computational thinking among students eligible for admission to a STEM-focused university?}}
\end{itemize}

In the challenge of inequality, some authors have raised the question for other marginalized groups in STEM subjects related to computational thinking. 
\new{Related research indicates that race and socio-economic status are influential factors in the participation and progress within related topics such as programming, computing, and the sciences.}

\new{Regarding race,} \citet{shah2020racial} analyzed participation in computational physics classrooms in the USA using an intersectional approach. They found that the white male teens usually took the lead in the different assignments, while the girls and the black teens tended to keep a secondary or nonexistent role in the classes' activities. This tendency differed when the groups were primarily composed by women and people of color \cite{shah2020racial}. Another study evaluated black students' participation and involvement in upstate New York programming and 3D printing workshops. Utilizing a participant ethnographic approach, \citet{lachney2017computational} concluded that the inclusion of elements of African culture - in this case, cornrow braids - in programming classes could provide new opportunities for development in programming and artificial intelligence, empowering segregated communities and addressing problems and challenges that are specific to these groups with low representation in the field.

\new{Considering another aspect,} \newtext{prior research did not find differences in computational thinking between private and public school students in Spain \cite{roman2017cognitive}. However,} in Chile, there is a strong relationship between the type of school, socioeconomic status, and academic performance~\cite{bellei2009public}. The Chilean educational system is known to be highly segregated by socioeconomic status~\cite{valenzuela2014socioeconomic}, where low-income families tend to enroll their children in public schools, middle-income families send their kids to subsidized (private voucher) schools, and high-income families heavily prefer private schools~\cite{elacqua2006school}. Public school students tend to score lower in standard math and language tests than those enrolled in subsidized schools. In turn, the latter achieve lower scores than students from private schools. The schools' average math and language test scores correlate with their average student income~\cite{mcewan2008school}. Thus, revealing a pattern of systematic inequality regarding socioeconomic status and academic performance. \newtext{Researchers argue that this pattern results from an education guided by supply and demand principles, leading to high-income and low-income schools \cite{valenzuela2014socioeconomic}. }

\newtext{This segregation has consequences for 
students' academic life in Chile, especially in the distribution of access to higher education. Focusing on science, \citet{pacheco2021examining} conducted a multilevel analysis of the PISA test scores of 15-year-old students and observed significant gaps. Socioeconomic status correlates to science achievement, expectations, and self-efficacy. This relationship means that lower-income students have worse results in science tests, lower expectations, and lower self-efficacy. Other researchers have analyzed the results of the University Selection Test, a national test taken by senior high school students to apply for college. They found a consistent socioeconomic gap in the test results across the years, with lower-income students having worse results than higher-income students \cite{santelices2015acceso,munoz2013inequality}. Thus, lower-income students have fewer alternatives to choose a graduate degree, having limited access to the most selective universities and degrees \cite{hastings2013some}, including STEM-related careers. Considering this context of inequality in education, especially in STEM-related subjects, it is necessary to analyze the role of the type of school in the development of computational thinking in a socioeconomically segregated educational system, such as the Chilean one.} \new{Therefore, the last research question is:}

\begin{itemize}
    \item \new{\textbf{RQ-3: Do inequalities exist in computational thinking skills across  types of school (private or not) within a socioeconomically segregated educational system?}}
\end{itemize}

\section{Research Methods}
\label{sec:researchmethods} 
\subsection{Study Design}

\new{To investigate the research questions, the authors designed a cross-sectional study aimed at assessing the computational thinking skills of students who had recently been admitted to a Chilean STEM-focused university. In Chile, students seeking higher education opportunities must go through a centralized admissions system, where they select their preferred universities and programs. They are also required to take the national University Selection Test, which comprises two mandatory sections: Language and Communication, and Mathematics. Furthermore, students have the option to take elective exams in subjects such as Science, History, and Social Sciences. The admission process takes into account students' preferences, their test scores, and their high school grade averages. Admission decisions are based on a weighted combination of these factors and the available vacancies in each program.}

\new{Historically, the university where the study was conducted has admitted students who score above the 80th percentile in the University Selection Test. Like many STEM programs in the country, the majority of enrolled students identify as males.}

\new{Given that this study was carried out in Chile, it is well-suited to address both RQ1 and RQ3. The Chilean educational system does not currently incorporate mandatory training in computational thinking~\cite{Pereiro2022,brackmann2016computational} (RQ1) and is characterized by socioeconomic disparities~\cite{valenzuela2014socioeconomic} (RQ3). The decision to examine the computational thinking skills of students eligible for STEM-focused university programs makes the study's sample appropriate for addressing RQ2.}

\new{To address the research questions, the study considered two primary independent variables: self-reported gender and type of school. Additionally, the researchers decided to incorporate prior programming knowledge as a variable of interest due to previous studies that have indicated a relationship between this type of training, gender, and computational thinking \cite{Atmatzidou_2016,Lai_2022}.}

\new{The researchers extended invitations to 1,200 students who had recently enrolled in the university. \moved{The invitation entailed completing a computational thinking assessment and had a requisite of being done before the first day of classes.} This aspect was significant for addressing RQ2, which seeks to determine whether a gender gap exists among students of STEM-related programs right before they start their first-year. \moved{Participation in the test was voluntary, and it was conducted during the 2019 admission cycle, prior to the disruptions caused by the COVID-19 pandemic.}}

\subsection{Measurement of Computational Thinking and Independent Variables}

\newtext{In conducting their study in Chile, the researchers} considered two aspects when selecting a computational thinking assessment instrument. First, Spanish is the country's official language, and all incoming college students must pass a language test in Spanish for admission. Consequently, \newtext{the researchers had confidence that} all participants would be proficient Spanish speakers, but they needed to refrain from assuming any knowledge of other languages. Second, the Chilean educational curricula do not include mandatory programming or computational thinking courses. Thus, the authors prioritized instruments that do not \new{require} participants to comprehend pseudo-code or a specific programming tool.

After considering these aspects, the researchers decided to utilize the Computational Thinking Test (CTt). Through visual-spatial questions, the CTt evaluates \newtext{basic directions and sequences}, 
loops, conditionals, and functions~\cite{gonzalez2015computational_design_guidelines}. The 
CTt \newtext{is an online test that} works independently from any programming tool and does not require prior programming knowledge. Experts have previously validated this instrument in Spanish~\cite{gonzalez2015computational_design_guidelines}. In Spain, CTt has been previously employed to diagnose computational thinking skills among students.





\newtext{While the CTt was initially designed for students in 5th to 10th grades~\cite{roman2015test}, the researchers used a CTt version tailored to older teens. This version was facilitated directly by the instrument designers. }\newtext{Previous research has validated and applied it to similar populations. \citet{chan2021assessing} offered evidence to substantiate the instrument validity by employing the Rasch scalability, utilizing a cohort of 153 upper-secondary students from Singapore. Additionally, \citet{guggemos2023computational} distributed it among 202 upper-secondary students in German-speaking Switzerland, \new{whose mean age was 17.23 years.}}

The CTt measures the construct of ``ability to formulate and solve problems based on the fundamental concepts of computing and using the logic and syntax of programming languages: basic sequences, loops, iterations, conditionals, functions, and variables.''~\cite{gonzalez2015computational_design_guidelines} \newtext{The test comprises 32 multiple-choice questions, each presenting four alternatives, of which only one is correct. }The CTt aims to evaluate achievement levels across four areas: \newtext{basic directions and sequences} (4 questions), conditionals (12 questions), loops (8 questions), and functions (8 questions).
%
%
\newtext{Basic directions and sequences refer to the understanding of constructing simple sequences of commands (e.g., move forward, turn left) to solve problems. Conditionals are used for decision-making (e.g., if, else, else if); loops refer to the understanding of repeating instructions multiple times. Finally, functions refer to utilizing reusable tools or procedures to perform specific tasks and operations.} 

\newtext{Moreover, the researchers incorporated} 18 additional questions in relation to students' demographics and educational circumstances, \newtext{encompassing factors} such as gender, school type, and prior programming knowledge.

\newtext{\subsection{Study Execution and Analysis}}
\label{sec:methodssurvey}

A total of 622 individuals responded to the CTt within the timeframe, out of which 593 consented to allow their answers to be used in this research. The dropout rate was 4.21\%, resulting in the completion of 568 tests. During the data analysis phase, participants who failed to provide information regarding their school type were excluded (15 cases). Additionally, individuals who did not identify as either male or female (4 individuals) were excluded due to the small sample size, which would have hindered the ability to conduct a robust statistical analysis. Consequently, the researchers analyzed the responses of 549 individuals who identified as male or female.

\newtext{The adapted CTt version used in this study had a strong level of reliability, as evidenced by a Cronbach's alpha value of $\alpha = 0.87$, slightly surpassing the reliability score attained in previous investigations ($\alpha = 0.79$)~\cite{tsarava2022cognitive}.}

The study examined the level of computational thinking, as measured by the score obtained in the CTt, considering three independent variables: self-reported gender (male or female), type of school (private vs. non-private), and prior programming knowledge (yes or no). The CTt score did not exhibit a normal distribution (skewness = -1.04, kurtosis = 3.59), and its values are positive integers within the range of 6 to 32. Therefore, the researchers used \newtext{a linear regression of the log-transformed CTt score to model} 
its association with the three independent variables.\footnote{The researchers also conducted Poisson regression to model the relationship between the raw CTt score and the same independent variables. The results were qualitatively similar.} 

To further delve into the analysis, separate regressions were conducted for each component of the CTt (\newtext{basic directions and sequences}, conditionals, loops, and functions), utilizing the same independent variables. Considering the score distribution for each component,  \newtext{a linear regression was employed to model the log-transformed score of the functions component}. In contrast, logistic regressions were employed for the scores of \newtext{basic directions and sequences}, conditionals, and loops. In the latter three cases, the dependent variable was binary, taking a value of 1 if the score attained in the component was above the median and 0 otherwise.

\new{The level of difficulty for each question was assessed through expert evaluation. Two computer science professors individually rated each item based on their experience with novice programming learners. Among the 32 questions, 68\% received identical ratings from both professors. Questions that received disparate ratings were subject to discussion until a consensus was reached. After this process, 13 questions were categorized as ``easy'', 15 as ``moderate difficulty'', and 4 as ``hard'', considering the study's target population.}

\section{Results of the Cross-Sectional Study}
\label{sec:analysissurvey}

\new{The students achieved an average score of $M= 25.5$ ($SD = 5.45$). The distribution of correct answers by question difficulty was as follows: 85\% of the students answered the easy questions correctly, 77\% for the medium-level questions, and 72\% for the difficult questions.}


\new{These results suggest that the students demonstrated strong computational thinking skills, as they achieved a high overall score on the test (25 out of 32), indicating that their success was not solely due to correct answers on easy questions.}

The majority of the student sample identified as male (65.9\%). \newtext{This proportion is very similar to the gender distribution (66.6\% male) in the university where the study was conducted}. A minority of students attended public schools (16.6\%), while the majority graduated from subsidized schools (55.4\%), and the remaining students received education from private schools (28.0\%) (see Table \ref{tab:1_students_by_type_of_school_gender}). About one-third of the sample (34.1\%) reported having prior programming knowledge acquired through formal or informal training facilitated by instructors or self-guided learning.

\begin{table}[t]
\begin{center}
\caption{Students by type of school and gender}
\label{tab:1_students_by_type_of_school_gender}
\begin{tabular}{lrrr}
\toprule
& Female & Male & Total (\%) \\ \midrule
Public school  & 25  & 66  & 91 (16.6)  \\
Private subsidized school  & 103  & 201  & 304 (55.4) \\
Private school & 59 & 95 & 154 (28.0) \\ \midrule
Total & 187 & 362 & 549 \\
(\%) & (34.1) & (65.9) & \\
\bottomrule
\end{tabular}
\end{center}
\end{table}

In the analyses, the researchers grouped students from public and subsidized schools \newtext{into a non-private school category} since there were no significant differences in the level of computational thinking achieved by these two groups. With this categorization, the results show that 33\% of students from public and subsidized schools had prior programming knowledge, whereas this proportion rose to 38\% among students from private schools.

Additionally, women were underrepresented among those who possess prior programming knowledge. While women account for approximately one-third of individuals from both private and non-private schools (see Table \ref{tab:2_students_by_type_school_gender_knowledge} (2)), their proportion is lower by 8 to 11 percentage points among individuals who declare having prior programming knowledge (see Table 2 (3b)). Specifically, while 32\% of students from public and subsidized schools who took the test are women, only 21\% of those with prior programming knowledge are women. This percentage decreases from 36\% to 28\% among private school students. Conversely, women are overrepresented among those without prior programming knowledge. Their proportion is between 6 and 9 percentage points higher among those who report that taking the CTt is their first exposure to programming than their prevalence in the whole sample (Table \ref{tab:2_students_by_type_school_gender_knowledge} (3a)).

\begin{table*}
\centering
\caption{Students by type of school, gender, and prior programming knowledge}
\label{tab:2_students_by_type_school_gender_knowledge}
\begin{tabular}{@{}lrrrrrrrrrrr@{}}
\toprule
& \multicolumn{2}{c}{(1)} & & \multicolumn{2}{c}{(2)} & & \multicolumn{2}{c}{(3a)} &
 & \multicolumn{2}{c}{(3b)}\\ 
 & \multicolumn{2}{c}{Type of school} & & \multicolumn{2}{c}{Gender (\%)} & & \multicolumn{2}{c}{No prior knowledge (\%)} &
 & \multicolumn{2}{c}{With prior knowledge (\%)}\\ 
 \cmidrule{2-3} \cmidrule{5-6}
\cmidrule{8-9} \cmidrule{11-12}
&& & & F & M && F & M && F & M  \\ 
 
Non-private school  && 395 &&    128 (32)  &  267 (68)&  &  100 (38) & 165 (62)      & & 28 (21)     & 102 (79)           \\
\midrule
Private school  && 154 &&    59 (36)  & 95 (64)    & & 43 (45)   & 53 (55)   & &  16 (28)  &     42 (72)        \\
\bottomrule
\end{tabular}
\end{table*}
\newtext{The researchers employed a linear regression }to estimate the relationship between the \newtext{log-transformed level of} computational thinking and the independent variables. Controlling for other variables, the results indicate (see Table \ref{tab:3_results_regression} (1)) that:

\begin{itemize}
    \item males attained CTt scores that were 1.1\newtext{3} times higher than those obtained by females ($p\leq0.001$),
    \item those who studied in private schools achieved scores that were 1.0\newtext{8} times higher than those obtained by students from non-private schools ($p=0.01$), and
    \item students with some prior programming knowledge reached scores that are 1.08 times higher than those obtained by students for whom the application of the CTt was their first approach to programming \newtext{($p=0.01$)}.
\end{itemize}

\begin{table*}
\def\sym#1{\ifmmode^{#1}\else\(^{#1}\)\fi}
\centering
\caption{Results of regressions for CTt scores and its components}
\label{tab:3_results_regression}
\begin{tabular}{llllll}
\toprule
\textbf{}   & \multicolumn{1}{c}{(1)} & \multicolumn{1}{c}{(2)} & \multicolumn{1}{c}{(3)}   & \multicolumn{1}{c}{(4)}   & \multicolumn{1}{c}{(5)}       \\
            & \multicolumn{1}{c}{CTt} & \multicolumn{1}{c}{\torevise{Directions and sequences}
            } & \multicolumn{1}{c}{Conditionals} & \multicolumn{1}{c}{Loops}    &\multicolumn{1}{c}{Functions} \\ \hline
Intercept          & \torevise{21.81}\sym{***}    & \textcolor{gray}{2.57}      & 0.49\sym{***}     & 1.15          & \torevise{5.57}\sym{***}\\
Gender (male)  & \torevise{1.13}\sym{***}     & \textcolor{gray}{1.46}      & 3.06\sym{***}     & 1.65\sym{*}   & \torevise{1.17}\sym{***}\\
Type of school (private)                   & \torevise{1.08}\sym{**}  & \textcolor{gray}{1.46}      & 1.48          & 1.59\sym{*}   & \torevise{1.12}\sym{*}\\
With prior programming knowledge   & \torevise{1.08}\sym{**} & \textcolor{gray}{1.34}      & 2.18\sym{***} & 1.95\sym{**}  & \torevise{1.10}\sym{*}\\ 
\midrule
Type of regression   & \torevise{Linear}           & \textcolor{gray}{Logistic}  & Logistic          & Logistic      & \torevise{Linear}   \\
p                   &  \torevise{3.91e-11}
& \textcolor{gray}{0.06187}   & 4.089e-13         & 1.967e-05     & \torevise{1.303e-05}\\
\bottomrule
\multicolumn{5}{l}{\footnotesize \sym{*} \(p<0.05\), \sym{**} \(p<0.01\), \sym{***} \(p<0.001\)}\\
\end{tabular}
\end{table*}

\begin{figure}[!htpb]
\centerline{\includegraphics[width=\linewidth]{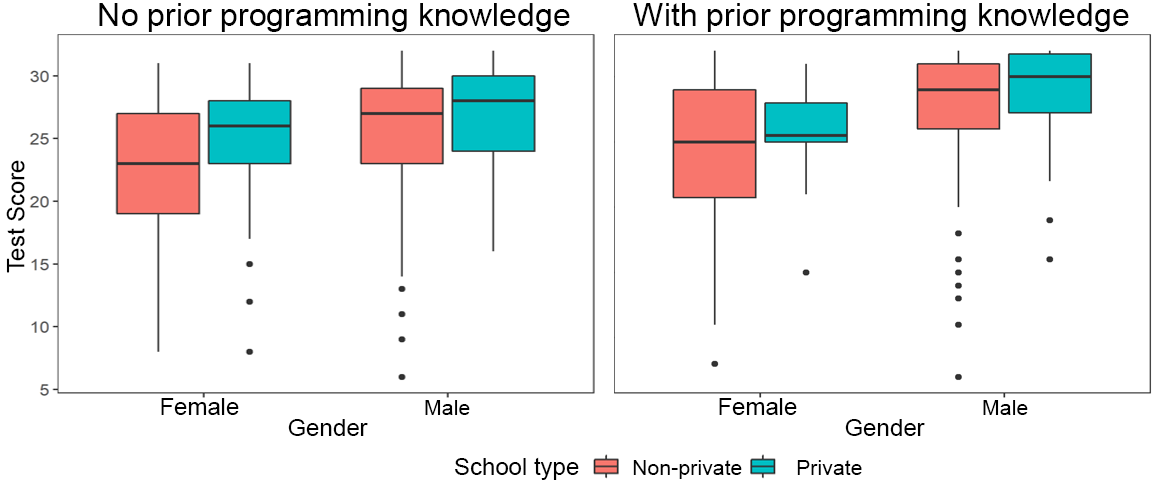}}
\caption{CTt score by gender, type of school, and prior programming knowledge in the cross-sectional study}
\label{fig:1}
\end{figure}

To illustrate these statistically significant differences, Figure \ref{fig:1} shows the distribution of the results obtained by students regarding the three independent variables. 

To further examine the effects of gender, type of school, and prior programming knowledge, four additional regressions were conducted to assess the significance of these differences concerning the four CTt components: \newtext{basic directions and sequences}, 
conditionals, loops, and functions (see Table \ref{tab:3_results_regression} (2-5)). The results indicate that the variables of gender, type of school, and prior programming knowledge do not predict achievement levels in the \newtext{basic directions and sequences} 
component ($p=0.06$). 
However, significant inequalities appeared in the other components. Gender and prior programming knowledge play a role in the conditionals, loops, and functions score. Inequalities related to the type of school are evident in only two components: loops and functions, which can be considered the more complex aspects of computational thinking measured by the CTt \cite{Gomes_2019, Cakiroglu_2022abstracion}.

\section{Discussion}
\label{sec:discussion}

\new{This work examined a sample of incoming students at a STEM university in Chile, and identified a robust development of computational thinking skills as evidenced by their ability to answer the majority of questions in a Computational Thinking Test, including those classified as highly challenging.}
This result indicates that despite the current Chilean school curriculum's limited emphasis on computational thinking, those pursuing STEM degrees demonstrate significant proficiency in these skills. \newtext{Thus, this research contributes valuable evidence about computational thinking development within a particular population in Latin America, which still needs to be studied more, given the historical US-centric focus \cite{saqr2021people} of computational thinking research. Further work could extend this research to encompass a broader geographical scope within Latin America, comparing the computational thinking abilities of students from different countries to uncover potential variations and common trends.}


However, it is essential to consider that this finding pertains specifically to the context of STEM students and may not be indicative of the broader population enrolled in Chilean universities or secondary education graduates. A plausible hypothesis could be that the high level of computational thinking results from the sample's composition, primarily comprising individuals with a keen interest in STEM and outstanding performance in the Chilean university admission test (above the 80th percentile). Future research could evaluate CTt among incoming students at other universities or those completing secondary education \newtext{(regardless of their college acceptance status)} to further investigate this possibility. 

The study also reveals significant gender disparities in computational thinking among students recently admitted into STEM undergraduate programs in Chile. In this context, female students demonstrate lower CTt scores than male peers. Interestingly, this discrepancy was absent among Spanish students in grades 5 and 6 but manifested in the higher grades (7 to 10)~\cite{gonzalez2015test}. \new{Prior work had also observed gender gaps in high-school students in Singapore \cite{chan2021assessing} and the UK \cite{Lai_2022}.} The current finding implies that gender-based inequities in computational thinking persist throughout secondary education, even among those who meet the qualifications for acceptance into STEM undergraduate degrees. This observation supports the hypothesis that the gender gap becomes apparent and endures as individuals age~\cite{roman2017cognitive}, particularly until the conclusion of high school, in contexts where computational thinking training is not compulsory. \moved{Future work could conduct longitudinal studies to track the development of computational thinking skills over time. This approach could provide deeper insights into how these skills evolve and how gender gaps emerge. }

\new{The gender gap observed in computational thinking aligns with previous findings regarding gender disparities in mathematics within Chile's education system, adding a new layer of complexity to the issue of gender imbalances in education.} \moved{For instance, using the PISA test, \citet{bharadwaj2016gender} identified a significant correlation between gender and mathematics test results, with 15-year-old females achieving lower scores than their male counterparts. Another study, utilizing the national system of learning outcomes assessment tests, revealed disparities in mathematics performance among 10th-grade students, where males outperformed females \cite{vargas2022brechas}.} \new{Similar to computational thinking}, \moved{the gender gap in mathematics persists until college admission when students take the University Selection Test \cite{santelices2015acceso}.} \moved{These trends are consistent with arguments proposed by other researchers, who suggest that gender disparities stem from gender stereotypes, differential expectations for males and females, and societal constructs that contribute to a perceived gap between women, mathematics, and technology ~\cite{Atmatzidou_2016,mujer2016informe}. } \new{Future studies should aim to identify the factors contributing to gender disparities in computational thinking and determine whether they are akin to or distinct from those affecting gaps in mathematics, in order to develop effective strategies for addressing both disparities.}

The data also indicates a positive correlation between higher computational thinking levels and prior programming knowledge. However, a more intriguing observation emerges when examining the data in Table \ref{tab:2_students_by_type_school_gender_knowledge}: female students are underrepresented among those who report having prior programming knowledge. This finding suggests that female students have had fewer opportunities to acquire programming knowledge before entering college, whether due to personal choice or other factors. This observation underscores the importance of factoring in prior programming knowledge when conducting gender-related analyses in computational thinking, both in the present study and future research. Furthermore, this result strengthens the case for implementing initiatives to provide more programming and technological training opportunities specifically targeted to girls and young women. \newtext{New studies could explore the potential long-term effects of these initiatives in narrowing the gender gap in computational thinking.}


\newtext{The combined results reveal a gender gap in a country where computational thinking training is voluntary, and female students are underrepresented among those who possess prior programming knowledge before enrolling in STEM-related universities. Thus, relying on voluntary and isolated workshops, instead of mandatory training, appears to have contributed to generating gender inequalities. 
Prior work ~\cite{hutchins2017role,Atmatzidou_2016} shows that after certain types of training (e.g., Scratch, educational robotics), there are no gender disparities in computational thinking. Thus, hinting that training (even if it is not targeted to women only) has an equalizing effect on gender inequalities. Future work should continue seeking evidence of the impact of one-fits-all interventions to mitigate gender gaps. If these interventions prove effective, policymakers could further advocate for integrating computational thinking into national education frameworks, highlighting the potential benefits for students' future careers and their potential to address gender imbalances in the subject.}




Related to gender, not all prior evidence supports the presence of a gap in computational thinking skills. The current research offers an explanation for previous contrasting outcomes. 
While some prior studies have found gender inequalities in primary and secondary schools~\cite{Kong_Lai_2022,Lai_2022,roman2017cognitive,chan2021assessing}, others have found no significant such disparities among first-year undergraduate engineering students~\cite{diaz2020computational,ata2020analysis}. The present findings indicate that the gender gap may persist until the final year of high school and before students are admitted to STEM-related degrees. It is possible that the gap narrows during the initial courses of engineering degrees, which would explain the results in ~\cite{diaz2020computational,ata2020analysis}. Early engineering courses might have a similar impact as training~\cite{hutchins2017role,Atmatzidou_2016}, leading to improvements in women's computational thinking skills to the extent that they are no longer significantly different from men's skills.

The researchers also observed significant disparities concerning the type of school attended by students. Those from private schools demonstrated higher performance levels in the CTt compared to students from public and subsidized schools. While administering a similar test to Spanish students did not provide evidence of gaps across types of schools~\cite{roman2017cognitive}, the observed gaps are not unexpected in Chile. The country already exhibits notable performance gaps in math and language tests among students from different school types, with private school students achieving the highest scores~\cite{mcewan2008school}. The data from this study shed light on a new dimension of systematic inequalities. Computational thinking emerges as another axis where students from private schools are better equipped than their peers from public and subsidized schools. 

Given the strong correlation between average scores in math and language and students' socioeconomic status~\cite{mcewan2008school}, as well as the considerable segregation within the Chilean educational system based on this variable~\cite{bellei2009public}, it is essential for future research to explore the link between socioeconomic status and computational thinking. Such investigations can clarify the observed disparities among different types of schools. \new{As socioeconomic status is associated with other forms of digital divide~\cite{van2020digital}, it is plausible} that computational thinking may also be associated with socioeconomic status, with the type of school serving as a proxy indicator within the context of Chile. More studies are needed to understand this connection at the individual and collective levels (e.g., school or country comparisons). This thread of research should delve deeper into the influence of socioeconomic factors on computational thinking skills, analyzing how broad (or limited) access to technology, internet connectivity, and other resources might contribute to inequalities in computational thinking.  


Together, the findings reveal a complex picture of the development of computational thinking in Chile. While the Chilean Ministry of Education has taken measures to provide computer labs, internet access, and an introductory computer class in all schools' curricula~\cite{MinistryofEducation_2022}, inequalities persist in advancing computational thinking within this context. Consequently, the results support that guaranteeing access to technology alone is not a determining factor in students' computational thinking performance. As previously suggested in the literature, there is no direct correlation between the skills required to operate computers and the cognitive traits associated with computational thinking~\cite{gibson20219}. Thus, further efforts are necessary to ensure equitable training in computational thinking, encompassing considerations of gender and school type.

Like any research, this work has limitations that warrant consideration. 
The CTt has certain constraints. Notably, it primarily emphasizes the conceptual aspect of computational thinking while only partially addressing computational practices and perspectives, which are also components of computational thinking. 
Furthermore, it is essential to acknowledge that the study participants were not randomly selected. Thus, our analyses do not account for all pre-existing factors that could have influenced the results. Factors related to self-selection could also explain some observed patterns.

Despite the acknowledged limitations, \newtext{the researchers anticipate their findings will have significant implications for future studies and potential policy changes in Chile, Latin America, and beyond. Specific relevant characteristics of the Chilean context are not unique to the country, such as high socioeconomic segregation within the educational system and a policy focus on technology access, leaving computational thinking development primarily to voluntary initiatives. Countries sharing similar characteristics could benefit from conducting analogous studies. }\newtext{The results concerning gender and school-type inequalities offer valuable insights to inspire such studies and inform discussions surrounding national plans to promote equitable access to programming. By seeking to democratize the efforts made by isolated and voluntary initiatives in computational thinking development, these countries can explore avenues to foster equal opportunities to choose and succeed in STEM-related degrees. For instance, implementing mandatory training may play a pivotal role in leveling the playing field for students of different genders and attending different types of schools within socioeconomically segregated educational systems.}


\section{Conclusion}
\label{sec:conclusion}

\new{This paper presents the findings of a cross-sectional study involving over 500 students admitted to STEM undergraduate programs at a Chilean university. The study highlights two significant gaps in the development of computational thinking skills: one related to gender and the other one associated with the type of school (private or not).}

\new{The results underscore the distinct characteristics in the development of computational thinking skills, particularly in regions like Latin America, where formal training in this topic is not mandatory and educational disparities are prevalent.} \new{These findings should be considered when expanding computational thinking training to address inequalities, as they offer valuable insights for countries aiming to promote equal access to programming, especially in socioeconomically segregated educational systems.}

\section*{Acknowledgments}
We thank Marcos Román-González, professor at Universidad Nacional de Educación a Distancia, for providing the ``Computational Thinking Test''. We also thank Dr. Federico Meza, Almendra Aguilera, and Dr. Pamela Soto for their support in different tasks of this project. Dr. Vasquez acknowledges the support of the UTFSM project OEA22202. Dr. López thanks the support of the grant PI\_L\_17\_12 from UTFSM, the Centro Nacional de Inteligencia Artificial CENIA, FB210017, BASAL, ANID, and the Millenium Nucleus Futures of AI and its socio-cultural implications in Chile and Latin America, FAIR, NCS2022\_065, ANID.


%

\bibliography{references}

\begin{thebibliography}{57}
\providecommand{\natexlab}[1]{#1}
\providecommand{\url}[1]{#1}
\csname url@samestyle\endcsname
\providecommand{\newblock}{\relax}
\providecommand{\bibinfo}[2]{#2}
\providecommand{\BIBentrySTDinterwordspacing}{\spaceskip=0pt\relax}
\providecommand{\BIBentryALTinterwordstretchfactor}{4}
\providecommand{\BIBentryALTinterwordspacing}{\spaceskip=\fontdimen2\font plus
\BIBentryALTinterwordstretchfactor\fontdimen3\font minus
  \fontdimen4\font\relax}
\providecommand{\BIBforeignlanguage}[2]{{%
\expandafter\ifx\csname l@#1\endcsname\relax
\typeout{** WARNING: IEEEtranN.bst: No hyphenation pattern has been}%
\typeout{** loaded for the language `#1'. Using the pattern for}%
\typeout{** the default language instead.}%
\else
\language=\csname l@#1\endcsname
\fi
#2}}
\providecommand{\BIBdecl}{\relax}
\BIBdecl

\bibitem[Wing(2006)]{wing_2006}
J.~M. Wing, ``Computational thinking,'' \emph{Commun. ACM}, vol.~49, no.~3, p.
  33–35, 2006.

\bibitem[Barr et~al.(2011)Barr, Harrison, and Conery]{barr2011computational}
D.~Barr, J.~Harrison, and L.~Conery, ``Computational thinking: A digital age
  skill for everyone.'' \emph{Learning \& Leading with Technology}, vol.~38,
  no.~6, pp. 20--23, 2011.

\bibitem[Monteiro and Leite(2021)]{Monteiro_Leite_2021}
A.~Monteiro and C.~Leite, ``Digital literacies in higher education: skills,
  uses, opportunities and obstacles to digital transformation,'' \emph{Revista
  de Educación a Distancia (RED)}, vol.~21, no.~65, 2021.

\bibitem[Enríquez et~al.(2016)Enríquez, Aguilar, and
  Domínguez]{enriquez_2016}
C.~Enríquez, O.~Aguilar, and F.~Domínguez, ``Using robot to motivate
  computational thinking in high school students,'' \emph{IEEE Latin America
  Transactions}, vol.~14, no.~11, pp. 4620--4625, 2016.

\bibitem[Brackmann et~al.(2017)Brackmann, Rom\'{a}n-Gonz\'{a}lez, Robles,
  Moreno-Le\'{o}n, Casali, and Barone]{Brackmann_2017_development}
C.~P. Brackmann, M.~Rom\'{a}n-Gonz\'{a}lez, G.~Robles, J.~Moreno-Le\'{o}n,
  A.~Casali, and D.~Barone, ``Development of computational thinking skills
  through unplugged activities in primary school,'' in \emph{Proceedings of the
  12th Workshop on Primary and Secondary Computing Education ({WiPSCE})}.\hskip
  1em plus 0.5em minus 0.4em\relax ACM, 2017, p. 65–72.

\bibitem[Heintz et~al.(2016)Heintz, Mannila, and Färnqvist]{heintz_2016}
F.~Heintz, L.~Mannila, and T.~Färnqvist, ``A review of models for introducing
  computational thinking, computer science and computing in k-12 education,''
  in \emph{2016 IEEE Frontiers in Education Conference (FIE)}, 2016, pp. 1--9.

\bibitem[Bocconi et~al.(2016)Bocconi, Chioccariello, Dettori, Ferrari,
  Engelhardt, et~al.]{bocconi2016developing}
S.~Bocconi, A.~Chioccariello, G.~Dettori, A.~Ferrari, K.~Engelhardt
  \emph{et~al.}, ``Developing computational thinking in compulsory
  education-implications for policy and practice,'' Joint Research Centre,
  Tech. Rep., 2016.

\bibitem[So et~al.(2020)So, Jong, and Liu]{so2020computational}
H.-J. So, M.~S.-Y. Jong, and C.-C. Liu, ``Computational thinking education in
  the asian pacific region,'' \emph{The Asia-Pacific Education Researcher},
  vol.~29, no.~1, pp. 1--8, 2020.

\bibitem[Moreno-León et~al.(2018)Moreno-León, Román-González, and
  Robles]{moreno_2017_oncomputational}
J.~Moreno-León, M.~Román-González, and G.~Robles, ``On computational
  thinking as a universal skill: A review of the latest research on this
  ability,'' in \emph{2018 IEEE Global Engineering Education Conference
  (EDUCON)}, 2018, pp. 1684--1689.

\bibitem[Lye and Koh(2014)]{Lye_2014}
S.~Y. Lye and J.~H.~L. Koh, ``Review on teaching and learning of computational
  thinking through programming: What is next for k-12?'' \emph{Computers in
  Human Behavior}, vol.~41, pp. 51--61, 2014.

\bibitem[Ericson and Guzdial(2014)]{Ericson_2014_Measuring}
B.~Ericson and M.~Guzdial, ``Measuring demographics and performance in computer
  science education at a nationwide scale using ap cs data,'' in
  \emph{Proceedings of the 45th ACM Technical Symposium on Computer Science
  Education ({SIGCSE})}.\hskip 1em plus 0.5em minus 0.4em\relax {ACM}, 2014, p.
  217–222.

\bibitem[Pereiro et~al.(2022)Pereiro, Montaldo, Victor, and
  Urruticoechea]{Pereiro2022}
E.~Pereiro, M.~Montaldo, K.~Victor, and A.~Urruticoechea, ``Computational
  thinking, artificial intelligence and education in {L}atin {A}merica,''
  {UNESCO}, Tech. Rep., 2022.

\bibitem[Simmonds et~al.(2021)Simmonds, Gutierrez, Meza, Torrent, and
  Villalobos]{Simmonds2021}
J.~Simmonds, F.~J. Gutierrez, F.~Meza, C.~Torrent, and J.~Villalobos,
  ``Changing teacher perceptions about computational thinking in grades 1-6,
  through a national training program,'' in \emph{Proceedings of the 52nd ACM
  Technical Symposium on Computer Science Education ({SIGCSE})}.\hskip 1em plus
  0.5em minus 0.4em\relax {ACM}, 2021, p. 260–266.

\bibitem[Lacoa et~al.(2016)Lacoa, Lacoa, and Blair]{lacoa_2016}
R.~F. Lacoa, J.~F. Lacoa, and A.~Blair, ``La ense{\~n}anza de lenguajes de
  programaci{\'o}n en la escuela: ¿por qu{\'e} hay que prestarle
  atenci{\'o}n?'' Fundaci{\'o}n Telef{\'o}nica, Tech. Rep., 2016.

\bibitem[Roig-Vila and Moreno-Isac(2020)]{Roig-Vila_Moreno-Isac_2020}
R.~Roig-Vila and V.~Moreno-Isac, ``El pensamiento computacional en educación.
  análisis bibliométrico y temático.'' \emph{Revista de Educación a
  Distancia}, vol.~20, no.~63, 2020.

\bibitem[Merino-Armero et~al.(2022)Merino-Armero, Gonz{\'a}lez-Calero, and
  Cozar-Gutierrez]{merino2022computational}
J.~M. Merino-Armero, J.~A. Gonz{\'a}lez-Calero, and R.~Cozar-Gutierrez,
  ``Computational thinking in k-12 education. an insight through
  meta-analysis,'' \emph{Journal of Research on Technology in Education},
  vol.~54, no.~3, pp. 410--437, 2022.

\bibitem[Rom{\'a}n-Gonz{\'a}lez et~al.(2017)Rom{\'a}n-Gonz{\'a}lez,
  P{\'e}rez-Gonz{\'a}lez, and Jim{\'e}nez-Fern{\'a}ndez]{roman2017cognitive}
M.~Rom{\'a}n-Gonz{\'a}lez, J.-C. P{\'e}rez-Gonz{\'a}lez, and
  C.~Jim{\'e}nez-Fern{\'a}ndez, ``Which cognitive abilities underlie
  computational thinking? {C}riterion validity of the computational thinking
  test,'' \emph{Computers in Human Behavior}, vol.~72, pp. 678--691, 2017.

\bibitem[Hutchins et~al.(2017)Hutchins, Zhang, and Biswas]{hutchins2017role}
N.~M. Hutchins, N.~Zhang, and G.~Biswas, ``The role gender differences in
  computational thinking confidence levels plays in stem applications,'' in
  \emph{The Int. Conference on Computational Thinking Education}, 2017, pp.
  34--38.

\bibitem[Atmatzidou and Demetriadis(2016)]{Atmatzidou_2016}
S.~Atmatzidou and S.~Demetriadis, ``Advancing students’ computational
  thinking skills through educational robotics: A study on age and gender
  relevant differences,'' \emph{Robotics and Autonomous Systems}, vol.~75, pp.
  661--670, 2016.

\bibitem[Mendoza~Diaz et~al.(2020)Mendoza~Diaz, Meier, Trytten, and
  Yoon]{diaz2020computational}
N.~V. Mendoza~Diaz, R.~Meier, D.~A. Trytten, and S.~Y. Yoon, ``Computational
  thinking growth during a first-year engineering course,'' in \emph{2020 IEEE
  Frontiers in Education Conference (FIE)}.\hskip 1em plus 0.5em minus
  0.4em\relax IEEE, 2020, pp. 1--7.

\bibitem[Ata and Y{\i}ld{\i}r{\i}m(2020)]{ata2020analysis}
R.~Ata and K.~Y{\i}ld{\i}r{\i}m, ``Analysis of the relation between
  computational thinking and new media literacy skills of first-year
  engineering students,'' \emph{Journal of Educational Multimedia and
  Hypermedia}, vol.~29, no.~1, pp. 5--20, 2020.

\bibitem[Lewis et~al.(2019)Lewis, Shah, and Falkner]{lewis201916}
C.~M. Lewis, N.~Shah, and K.~Falkner, ``16 equity and diversity,'' \emph{The
  Cambridge handbook of computing education research}, p. 481, 2019.

\bibitem[Shah et~al.(2020)Shah, Christensen, Ortiz, Nguyen, Byun, Stroupe, and
  Reinholz]{shah2020racial}
N.~Shah, J.~A. Christensen, N.~A. Ortiz, A.-K. Nguyen, S.~Byun, D.~Stroupe, and
  D.~L. Reinholz, ``Racial hierarchy and masculine space: Participatory
  in/equity in computational physics classrooms,'' \emph{Computer Science
  Education}, vol.~30, no.~3, pp. 254--278, 2020.

\bibitem[Lei et~al.(2020)Lei, Chiu, Li, Wang, and Geng]{lei2020computational}
H.~Lei, M.~M. Chiu, F.~Li, X.~Wang, and Y.-j. Geng, ``Computational thinking
  and academic achievement: A meta-analysis among students,'' \emph{Children
  and Youth Services Review}, vol. 118, p. 105439, 2020.

\bibitem[Garc{\'\i}a-Pe{\~n}alvo and Mendes(2018)]{garcia2018exploring}
F.~J. Garc{\'\i}a-Pe{\~n}alvo and A.~J. Mendes, ``Exploring the computational
  thinking effects in pre-university education,'' \emph{Computers in Human
  Behavior}, vol.~80, pp. 407--411, 2018.

\bibitem[Quiroz-Vallejo et~al.(2021)Quiroz-Vallejo, Carmona-Mesa,
  Castrill{\'o}n-Yepes, and Villa-Ochoa]{quiroz2021integration}
D.~A. Quiroz-Vallejo, J.~A. Carmona-Mesa, A.~Castrill{\'o}n-Yepes, and J.~A.
  Villa-Ochoa, ``Integration of computational thinking in elementary and
  secondary school in latin america: A systematic literature,'' \emph{Revista
  de Educación a Distancia (RED)}, vol.~21, no.~68, 2021.

\bibitem[Brackmann et~al.(2016)Brackmann, Barone, Casali, Boucinha, and
  Mu{\~n}oz-Hernandez]{brackmann2016computational}
C.~Brackmann, D.~Barone, A.~Casali, R.~Boucinha, and S.~Mu{\~n}oz-Hernandez,
  ``Computational thinking: Panorama of the americas,'' in \emph{{International
  Symposium on Computers in Education}}.\hskip 1em plus 0.5em minus 0.4em\relax
  IEEE, 2016, pp. 1--6.

\bibitem[Age(2015)]{AgendaDigital_2020}
``Agenda digital. {Imagina Chile} 2013 - 2020,'' Gobierno de Chile; Secretaría
  Ejecutiva de Desarrollo, Santiago, Chile, Tech. Rep., 2015.

\bibitem[Peroni~Fiscarelli et~al.(2018)Peroni~Fiscarelli, Escobar~Riffo, and
  Escobedo~Seguel]{fiscarelli2018informe}
A.~Peroni~Fiscarelli, D.~Escobar~Riffo, and C.~Escobedo~Seguel, ``Informe final
  de evaluaci{\'o}n programs gubernamentales {(EPG)},'' 2018.

\bibitem[Min(2014)]{MinistryofEducation_2014}
``{SIMCE TIC} 2013, {E}valuación de habilidades tic para el aprendizaje.
  resultados nacionales,'' Enlaces, Centro de Educación y Tecnología.
  Ministerio de Educación, Santiago, Chile, Tech. Rep., 2014.

\bibitem[Lai(2022)]{Lai_2022}
R.~P.~Y. Lai, ``\BIBforeignlanguage{en}{Beyond programming: A computer-based
  assessment of computational thinking competency},''
  \emph{\BIBforeignlanguage{en}{ACM Transactions on Computing Education}},
  vol.~22, no.~2, p. 1–27, 2022.

\bibitem[Chan et~al.(2021)Chan, Looi, and Sumintono]{chan2021assessing}
S.-W. Chan, C.-K. Looi, and B.~Sumintono, ``Assessing computational thinking
  abilities among singapore secondary students: a rasch model measurement
  analysis,'' \emph{Journal of Computers in Education}, vol.~8, pp. 213--236,
  2021.

\bibitem[Kong and Lai(2022)]{Kong_Lai_2022}
S.-C. Kong and M.~Lai, ``\BIBforeignlanguage{en}{Validating a computational
  thinking concepts test for primary education using item response theory: An
  analysis of students’ responses},'' \emph{\BIBforeignlanguage{en}{Computers
  \& Education}}, vol. 187, p. 104562, 2022.

\bibitem[Sullivan and Bers(2016)]{sullivan2016girls}
A.~Sullivan and M.~U. Bers, ``Girls, boys, and bots: Gender differences in
  young children’s performance on robotics and programming tasks,''
  \emph{Journal of Information Technology Education. Innovations in Practice},
  vol.~15, p. 145, 2016.

\bibitem[Lachney(2017)]{lachney2017computational}
M.~Lachney, ``Computational communities: African-american cultural capital in
  computer science education,'' \emph{Computer Science Education}, vol.~27, no.
  3-4, pp. 175--196, 2017.

\bibitem[Bellei(2009)]{bellei2009public}
C.~Bellei, ``The public-private school controversy in {C}hile,'' in
  \emph{School Choice Internacional. Exploring Public-Private Partnerships},
  R.~Chakrabarti and P.~E. Peterson, Eds.\hskip 1em plus 0.5em minus
  0.4em\relax The MIT Press, 2009, ch.~8, pp. 165--192.

\bibitem[Valenzuela et~al.(2014)Valenzuela, Bellei, and de~los
  R{\'\i}os]{valenzuela2014socioeconomic}
J.~P. Valenzuela, C.~Bellei, and D.~de~los R{\'\i}os, ``{Socioeconomic school
  segregation in a market-oriented educational system. The case of Chile},''
  \emph{Journal of Education Policy}, vol.~29, no.~2, pp. 217--241, 2014.

\bibitem[Elacqua et~al.(2006)Elacqua, Schneider, and
  Buckley]{elacqua2006school}
G.~Elacqua, M.~Schneider, and J.~Buckley, ``School choice in {C}hile: Is it
  class or the classroom?'' \emph{Journal of Policy Analysis and Management},
  vol.~25, no.~3, pp. 577--601, 2006.

\bibitem[McEwan et~al.(2008)McEwan, Urquiola, Vegas, Fernandes, and
  Gallego]{mcewan2008school}
P.~J. McEwan, M.~Urquiola, E.~Vegas, R.~Fernandes, and F.~A. Gallego, ``School
  choice, stratification, and information on school performance: Lessons from
  {C}hile,'' \emph{Economia}, vol.~8, no.~2, pp. 1--42, 2008.

\bibitem[Diaz and Rocconi(2021)]{pacheco2021examining}
N.~P. Diaz and L.~Rocconi, ``Examining science achievement in {C}hile: A
  multilevel model approach,'' \emph{Journal of Research in STEM Education},
  vol.~7, no.~2, pp. 93--116, 2021.

\bibitem[Santelices et~al.(2015)Santelices, Galleguillos, and
  Catal{\'a}n]{santelices2015acceso}
M.~V. Santelices, P.~Galleguillos, and X.~Catal{\'a}n, ``El acceso y la
  transici{\'o}n a la universidad en {Chile},'' \emph{A. Bernasconi, La
  educaci{\'o}n superior en {C}hile: transformaci{\'o}n, desarrollo y crisis.
  Santiago de {C}hile: Colecci{\'o}n Educaci{\'o}n Superior, Ediciones {UC}},
  2015.

\bibitem[Mu{\~n}oz and Redondo(2013)]{munoz2013inequality}
P.~Mu{\~n}oz and A.~Redondo, ``Inequality and academic achievement in
  {C}hile,'' \emph{Cepal Review}, no. 109, 2013.

\bibitem[Hastings et~al.(2013)Hastings, Neilson, and
  Zimmerman]{hastings2013some}
J.~S. Hastings, C.~A. Neilson, and S.~D. Zimmerman, ``Are some degrees worth
  more than others? evidence from college admission cutoffs in {C}hile,''
  National Bureau of Economic Research, Tech. Rep., 2013.

\bibitem[Gonz{\'a}lez(2015{\natexlab{a}})]{gonzalez2015computational_design_guidelines}
M.~R. Gonz{\'a}lez, ``Computational thinking test: Design guidelines and
  content validation,'' in \emph{Proceedings of EDULEARN15 conference}, 2015,
  pp. 2436--2444.

\bibitem[Rom{\'a}n-Gonzalez et~al.(2015)Rom{\'a}n-Gonzalez,
  P{\'e}rez-Gonz{\'a}lez, and Jim{\'e}nez-Fern{\'a}ndez]{roman2015test}
M.~Rom{\'a}n-Gonzalez, J.~C. P{\'e}rez-Gonz{\'a}lez, and
  C.~Jim{\'e}nez-Fern{\'a}ndez, ``Test de pensamiento computacional: dise{\~n}o
  y psicometr{\'\i}a general,'' in \emph{{III} {Congreso Internacional sobre
  Aprendizaje, Innovaci{\'o}n y Competitividad (CINAIC)}}, 2015, pp. 1--6.

\bibitem[Guggemos et~al.(2023)Guggemos, Seufert, and
  Rom{\'a}n-Gonz{\'a}lez]{guggemos2023computational}
J.~Guggemos, S.~Seufert, and M.~Rom{\'a}n-Gonz{\'a}lez, ``Computational
  thinking assessment--towards more vivid interpretations,'' \emph{Technology,
  Knowledge and learning}, vol.~28, no.~2, pp. 539--568, 2023.

\bibitem[Tsarava et~al.(2022)Tsarava, Moeller, Rom{\'a}n-Gonz{\'a}lez, Golle,
  Leifheit, Butz, and Ninaus]{tsarava2022cognitive}
K.~Tsarava, K.~Moeller, M.~Rom{\'a}n-Gonz{\'a}lez, J.~Golle, L.~Leifheit, M.~V.
  Butz, and M.~Ninaus, ``A cognitive definition of computational thinking in
  primary education,'' \emph{Computers \& Education}, vol. 179, p. 104425,
  2022.

\bibitem[Gomes et~al.(2019)Gomes, Ke, Lam, Teixeira, Correia, Marcelino, and
  Mendes]{Gomes_2019}
A.~Gomes, W.~Ke, C.-T. Lam, A.~R. Teixeira, F.~B. Correia, M.~J. Marcelino, and
  A.~J. Mendes, ``Understanding loops: a visual methodology,'' in \emph{2019
  IEEE Int. Conf. on Engineering, Technology and Education}, 2019, p. 1–7.

\bibitem[\c{C}akiro\u{g}lu and \c{C}evik(2022)]{Cakiroglu_2022abstracion}
U.~\c{C}akiro\u{g}lu and I.~\c{C}evik, ``A framework for measuring abstraction
  as a sub-skill of computational thinking in block-based programming
  environments,'' \emph{Education and Information Technologies}, vol.~27, p.
  9455–9484, 2022.

\bibitem[Saqr et~al.(2021)Saqr, Ng, Oyelere, and Tedre]{saqr2021people}
M.~Saqr, K.~Ng, S.~S. Oyelere, and M.~Tedre, ``People, ideas, milestones: a
  scientometric study of computational thinking,'' \emph{ACM Transactions on
  Computing Education (TOCE)}, vol.~21, no.~3, pp. 1--17, 2021.

\bibitem[Gonz{\'a}lez(2015{\natexlab{b}})]{gonzalez2015test}
M.~R. Gonz{\'a}lez, ``Test de pensamiento computacional: principios de
  dise{\~n}o, validaci{\'o}n de contenido y an{\'a}lisis de {\'\i}tems,'' in
  \emph{Perspectivas y avances de la investigaci{\'o}n: {I} Jornada de
  Doctorandos}, 2015, pp. 291--314.

\bibitem[Bharadwaj et~al.(2016)Bharadwaj, De~Giorgi, Hansen, and
  Neilson]{bharadwaj2016gender}
P.~Bharadwaj, G.~De~Giorgi, D.~Hansen, and C.~A. Neilson, ``The gender gap in
  mathematics: evidence from {C}hile,'' \emph{Economic Development and Cultural
  Change}, vol.~65, no.~1, pp. 141--166, 2016.

\bibitem[Vargas~Diaz and Matus~Correa(2022)]{vargas2022brechas}
C.~Vargas~Diaz and C.~Matus~Correa, ``Brechas persistentes de g{\'e}nero en
  matem{\'a}ticas en las pruebas nacionales chilenas simce,'' \emph{Estudios
  pedag{\'o}gicos (Valdivia)}, vol.~48, no.~1, pp. 389--400, 2022.

\bibitem[muj(2016)]{mujer2016informe}
``Informe {GET} (g{\'e}nero, educaci{\'o}n y trabajo). la brecha persistente.
  primer estudio sobre la desigualdad de g{\'e}nero en el ciclo de vida. una
  revisi{\'o}n de los {\'u}ltimos 25 a{\~n}os,'' Comunidad Mujer, Chile, Tech.
  Rep., 2016.

\bibitem[Van~Dijk(2020)]{van2020digital}
J.~Van~Dijk, \emph{The digital divide}.\hskip 1em plus 0.5em minus 0.4em\relax
  John Wiley \& Sons, 2020.

\bibitem[Min(2022)]{MinistryofEducation_2022}
``Historia centro de innovación en educación,'' Centro de Educación y
  Tecnología ENLACES, Ministerio de Educación 1, Santiago, Chile, Tech. Rep.,
  2022.

\bibitem[Gibson and Brown(2021)]{gibson20219}
P.~Gibson and M.~Brown, ``9 computational thinking and technology-enhanced
  learning {(TEL)},'' in \emph{Inspiring Primary Learners: Insights and
  Inspiration Across the Curriculum}, R.~McDonald and P.~Gibson, Eds.\hskip 1em
  plus 0.5em minus 0.4em\relax Routledge, 2021, ch.~9, pp. 151--168.

\end{thebibliography}
\bibliographystyle{IEEEtranN}

\vspace{-1.2cm} 
\begin{IEEEbiography}
{Felipe González-Pizarro}  holds a B.Sc. and an M.Sc. in Computer Science from Universidad Técnica Federico Santa María, Chile. He also holds an M.Sc. in Computer Science from the University of British Columbia, Canada. His research interests encompass natural language processing, multimodal machine learning, information visualization, and social computing. Additionally, he has actively participated in various initiatives aimed at promoting the development of computational thinking in high schools and universities. 
\end{IEEEbiography}
\vspace{-1.2cm} 
\begin{IEEEbiography}
{Claudia López} is an assistant professor in the Informatics Department at Universidad Técnica Federico Santa María, Chile. With a Ph.D. in Information Sciences and Technology obtained from the University of Pittsburgh, she works as an associate researcher at the Chilean National Center of Artificial Intelligence (CENIA) and as a principal investigator in the Millenium Nucleus Futures of AI (FAIR). Claudia advocates for gender equity in computing.
\end{IEEEbiography}
\vspace{-1.2cm} 
\begin{IEEEbiography}
{Andrea Vásquez}, a software engineer with a Ph.D. in engineering sciences from Pontificia Universidad Catolica de Chile, is dedicated to teaching programming at various levels. She develops educational materials and workshops to promote computational thinking learning and serves as a faculty member at Universidad Tecnica Federico Santa Maria. Andrea's research focuses on computer science education, educational technology, and computer-supported collaborative learning.
\end{IEEEbiography}
\vspace{-1.2cm} 
\begin{IEEEbiography}
{Carlos Castro} holds B.Sc. and M.Sc. degrees in Informatics from Universidad Técnica Federico Santa María, Chile, and a  Docteur en Informatique degree from the Université Henri Poincaré, France. 
He specializes in solving optimization problems using traditional operations research. Carlos is also dedicated to fostering educational development in computing.
\end{IEEEbiography}

\end{document}